\documentstyle[aps,prb,epsf,preprint]{revtex}

\begin{document}
\tightenlines
\title{Anomalous magnetic response of the spin-one-half Falicov-Kimball model}
\author{J. K. Freericks and V. Zlati\'c$^{\dag}$}
\address{Department of Physics, Georgetown University, Washington, DC 20057}
\date{\today }
\maketitle

\begin{abstract}
The infinite-dimensional spin one-half Falicov-Kimball model in an external
magnetic field is solved exactly.  We calculate the magnetic susceptibility
in zero field, and the magnetization as a function of the field strength.
The model shows an anomalous magnetic response from thermally excited
local moments that disappear as the temperature is lowered.  We describe
possible real materials that may exhibit this kind of anomalous behavior.
\end{abstract}


\draft

\renewcommand{\thefootnote}{\copyright}
\footnotetext{ 1998 by the authors.  Reproduction of this article by any means
is permitted for non-commercial purposes.}
\renewcommand{\thefootnote}{\alpha{footnote}}

\section{Introduction}

The spin-one-half Falicov-Kimball model \cite{falicov.69} was introduced
in 1969 to describe metal-insulator transitions in transition-metal and
rare-earth compounds.  The metal-insulator transition is a charge-transfer
transition, where the system consists of both localized (insulating) electronic
states and delocalized (conducting) electronic states, and the transition 
occurs when the electron filling switches from the localized to delocalized
states (or vice versa).  This is the simplest possible metal-insulator 
transition, since the character of the individual
electronic states does not change at the transition, rather it is just the
occupancy of those states that varies.  

The localized electrons are strongly interacting with each other, so
double occupancy of the localized orbitals is forbidden.  The conduction 
electrons, however, are chosen to be noninteracting, since the Coulomb 
interaction between them is screened.
The only remaining interaction term in the Falicov-Kimball model is the 
mutual repulsion between a conduction and localized electron that sit on the 
same lattice site.  For any fixed configuration of the localized electrons,
the quantum-mechanical problem for the conduction electrons can be solved
in a single-particle basis.  The many-body aspects enter
from taking an annealed thermal average 
over all possible configurations of the localized electrons.  

We are most
interested in the situation where the localized level lies just above the
Fermi energy, so there are no localized electrons at $T=0$.  As the
temperature rises, then it becomes more favorable to thermally occupy the 
localized levels because of the gain in entropy of the local moments.
This will have a dramatic effect on the uniform spin susceptibility,
as the localized electrons will generate a Curie-like response, which
is much larger than the Pauli-like response of the conduction electrons.
The net effect is that the susceptibility will show a peak at a characteristic
temperature which is determined by the point where the localized electron
occupancy saturates.  In addition, we expect there to be interesting
behavior as a function of magnetic field, where the proximity of the
localized levels close to the Fermi energy can induce a rapid change in
the magnetization, producing metamagnetism.

The Falicov-Kimball
model was initially applied to a wide variety of transition-metal and
rare-earth compounds such as SmB$_6$, V$_2$O$_3$, NiS, etc. which display both
continuous and discontinuous metal-insulator transitions as a function of
either temperature or pressure.  However, the Falicov-Kimball model has not
been generally accepted as explaining the metal-insulator transitions of 
all of these different materials for two reasons.  First, it neglects all
effects of hybridization, and hence all Kondo-effect physics, and second it
does not take into account the fact that most of these materials also
undergo a structural phase transition.

Recently, however, a new material was discovered that undergoes a classic
charge-transfer metal-insulator transition without any structural phase
transition---it is NiI$_2$ when put under high pressure\cite{nii2}.  This 
layered compound crystallizes in the CdCl$_2$ crystal structure, in which the 
Ni ions form a close-packed plane, with close-packed I planes lying 
both above and below. The NiI$_2$ planes are then stacked in an alternating
packing scheme. The
Fermi level of NiI$_2$ lies within the Ni d-bands, since the Ni donates one
electron to each I ion that lies in the plane above and
below.  The ground state is an antiferromagnetic insulator.
As pressure is applied, the I p-bands move closer to the Fermi level
causing the N\'eel temperature to rise, until they reach the Fermi 
level, and electrons from the I ions move onto the Ni$^{++}$ ions changing
them to Ni$^+$ and quenching the local magnetic moment.  The remaining holes
in the I p-bands are conducting, and the material becomes metallic.

Another class of materials that might be described by an effective 
Falicov-Kimball model is the class of
Yb-based valence-fluctuating (VF) compounds (like YbInCu$_4$, YbIn$%
_{1-x} $Ag$_x$Cu$_4$ and Yb$_{1-x}$Y$_x$InCu$_4$) which exhibit large 
anomalies in
their thermodynamic \cite{felner.86,yoshimura.88,kindler.94}, spectroscopic 
\cite{felner.86,kojima.90,nakamura.90} and transport properties \cite
{felner.86,kindler.94,sarrao.96}, around a characteristic, sample-dependent
temperature $T_V$ ($40-60K$). 
The magnetic susceptibility $\chi (T)$ of these compounds is  
of a Curie-Weiss form at high temperatures, and the anomaly 
appears\cite{kindler.94,sarrao.96,junhui.97}, as
a pronounced asymmetric peak just above this characteristic sample-dependent
temperature $T_V$. 
Below $T_V$, $\chi (T)$ exhibits Pauli-like behavior, while at the lowest
temperatures a weak Curie-like upturn is often seen. 
The X-ray analysis shows that the Yb-based VF compounds crystallize in 
a C15b structure, that the  lattice constant $a(T)$ changes at $T_V$ by a 
small amount \cite{kojima.90,sarrao.96,lawrence.96}, and that there are no
structural changes. There is no evidence\cite{yoshimura.88,kojima.90} for 
magnetic ordering below $T_V$, instead the 
Yb ions fluctuate between a $2+$ and $3+$
state, but the average volume change at $T_V$ is much less than what one 
would expect from a complete Yb$^{3+}$--Yb$^{2+}$ transition.\cite{felner.86}
The anomaly in the high-field magnetization of YbInCu$_4$\cite{yoshimura.88}
and YbIn$_{1-x}$Ag$_x$Cu$_4$\cite{katori.94} appears as a sudden increase of
the slope and the saturation of $M(H)$ at about $H_V\simeq 30-50\;T$. The
Zeeman energy at $H_V$ (at $T\ll T_V$) is comparable to the thermal energy
at $T_V$ (at $H\ll H_V$). At elevated temperatures, this metamagnetic
transition disappears\cite{kojima.90}. The magnetostriction data\cite
{yoshimura.88} indicate that the metamagnetism relates to the valence
fluctuations of the Yb ions but the field-induced change of the $f$-electron
valence is even smaller than the temperature-induced one.

The large magnetic anomalies accompanied by small lattice changes can
be reconciled by a model in which
there is some intrinsic disorder between Yb and In sites, 
such that a minority of Yb ions (in the C15b lattice) lie at  
``ill--placed sites''. The correctly placed
Yb ions are hybridized with the ligands and are in the VF state 
at all temperatures. The ``ill-placed'' Yb ions are unhybridized 
(because the Yb-In distance of an ``ill-placed'' Yb ion [that sits on
an In site] is nearly twice as large as the Yb-In distance for an Yb ion
that sits on the correct lattice site) and undergo 
a charge-transfer transition at $T_V$
(one 4f electron jumps into conduction band and leaves behind a magnetic hole). 
As a consequence, the susceptibility acquires a large asymmetric peak 
just above  $T_V$, while the low-temperature magnetization exhibits a 
metamagnetic transition for magnetic fields of the order of $k_BT_V$. 

This recent experimental work
provides our motivation to study the magnetic response of the spin-one-half
Falicov-Kimball model, and to compare the results with the available
susceptibility data for  NiI$_2$ and YbInCu$_4$. On the theoretical side,
the Falicov-Kimball model has also received renewed interest, with work
concentrating on the linear and nonlinear characteristics of the
model\cite{nonlinear} and on the intermediate-valence and metal-insulator
transitions\cite{farkasovsky}.

In Section II we present the formalism, showing how to determine the
Green's functions in a magnetic field, and the different magnetic and
charge-density-wave response functions in zero field, and the numerical
methods employed in the exact solution of the problem.  Section III presents
our results for the infinite-dimensional hypercubic lattice and Section IV
summarizes the possible applications to real materials.  Our conclusions appear
in Section V.

\section{Formalism}

The Hamiltonian of the Falicov-Kimball model consists of two types of
spin-one-half electrons:  conduction electrons (created or destroyed at site $i$
by $d_{i\sigma}^{\dagger}$ or $d_{i\sigma}$) and localized electrons 
(created or destroyed at site $i$ by $f_{i\sigma}^{\dagger}$ or $f_{i\sigma}$). 
The conduction electrons can hop 
between nearest-neighbor sites, with a hopping matrix $-t_{ij}=:-t^*/2\sqrt{D}$,
where we have chosen to examine hypercubic lattices in $D$-dimensions, and
we choose a scaling of the hopping matrix that yields a nontrivial limit
in infinite-dimensions ($D\rightarrow\infty$)
\cite{metzner_vollhardt}.  The $f$-electrons have
a site energy $E_f$ (which will be chosen to lie just above the Fermi
energy at $T=0$), and a chemical potential $\mu$ is employed to
conserve the total number of electrons $n_{d\uparrow}+n_{d\downarrow}+
n_{f\uparrow}+n_{f\downarrow}=~{\rm constant}$.  Finally, there is
a restriction of no double occupancy of the $f$-electrons on any site
(implying the Coulomb repulsion $U_{ff}$ between two $f$-electrons is infinite)
and there is a Coulomb interaction $U$ between the $d$- and $f$-electrons
that occupy the same lattice site.  In addition, we include a coupling to
an external magnetic field $h$ with a Land\'e g-factor added for the
localized electrons.  The resulting Hamiltonian is \cite{falicov.69,brandt.1989}
\begin{eqnarray}
H=\sum_{ij,\sigma }(-t_{ij}-\mu \delta _{ij})d_{i\sigma }^{\dagger
}d_{j\sigma }+\sum_{i,\sigma }(E_f-\mu )f_{i\sigma }^{\dagger }f_{i\sigma }%
\cr +U\sum_{i,\sigma \sigma ^{\prime }}d_{i\sigma }^{\dagger }d_{i\sigma
}f_{i\sigma ^{\prime }}^{\dagger }f_{i\sigma ^{\prime
}}+U_{ff}\sum_{i,\sigma }f_{i\uparrow }^{\dagger }f_{i\uparrow
}f_{i\downarrow }^{\dagger }f_{i\downarrow }\cr -\mu _Bh\sum_{i,\sigma
}\sigma (2d_{i\sigma }^{\dag }d_{i\sigma }+gf_{i\sigma }^{\dag }f_{i\sigma
}).  \label{eq:H_FK}
\end{eqnarray}

We solve the Falicov-Kimball model by working in the infinite-dimensional
limit.  The bare density of states $\rho(\epsilon)$ on a hypercubic lattice
becomes 
\begin{equation}
\rho(\epsilon)=\frac{1}{\sqrt{\pi}t^*}\exp [-\epsilon^2/t^{*2}],
\label{eq: dosdef}
\end{equation}
and we take $t^*$ as the unit of energy ($t^*=1$).  (The alternative lattice to
consider is the Bethe lattice, where the bare density of states becomes Wigner's
semicircle.  The latter density of states is closer to a three-dimensional
band because it increases like $\sqrt{E}$ at the band edges.  Since the 
Fermi level always lies far away from the band edge in all calculations
considered here, we don't expect there to be much difference between the
solution on a Bethe lattice versus on a hypercubic lattice.  In fact, we have
checked that this is not the case for our results in Section III.) In the 
infinite-dimensional limit the local approximation becomes 
exact\cite{metzner_vollhardt} so that the self energy has no momentum
dependence, and the local conduction-electron Green's function satisfies
\begin{equation}
G_{\sigma}(i\omega_n)=:G_{n\sigma}=\int \frac{\rho(\epsilon)}
{i\omega_n+\mu-\Sigma_{n\sigma}-\epsilon}d\epsilon,
\label{eq: gloc}
\end{equation}
where $\omega_n=:\pi T(2n+1)$ is the Fermionic Matsubara frequency,
and $\rho(\epsilon)$ is the noninteracting density of states.
The self-energy is determined by a three-step process:  (i) first,
the site-excluded Green's function $G_{\sigma}^0(i\omega_n)$, which 
holds the information about all of the other lattice sites, is found
by adding back the self energy,
\begin{equation}
G_{\sigma}^0(i\omega_n)=\frac{1}{G_{n\sigma}^{-1}+\Sigma_{n\sigma}};
\label{eq: g0def}
\end{equation}
then (ii) the local Green's function is calculated from a weighted sum over the 
different possible localized-electron states,
\begin{equation}
G_{n\sigma}=w_0G_{\sigma}^0(i\omega_n)+\frac{w_1^{\uparrow}+w_1^{\downarrow}}
{[G_{\sigma}^0(i\omega_n)]^{-1}-U},
\label{eq: gloc2}
\end{equation}
where $w_0=1-w_1^{\uparrow}-w_1^{\downarrow}$ and $w_1^{\sigma}$ are the
localized electron occupancies for no f-electrons ($w_0$) and an
f-electron with spin $\sigma$ ($w_1^{\sigma}$); and finally (iii) the
self-energy is determined by subtracting off the ``bare'' Green's function
\begin{equation}
\Sigma_{n\sigma}=[G_{\sigma}^0(i\omega_n)]^{-1}-G_{n\sigma}^{-1}.
\label{eq: self}
\end{equation}
The f-electron occupancies are found by explicitly calculating the 
path-integral for the partition function\cite{brandt.1989} (generalized
to include the magnetic field) and are determined by $w_0={\cal Z}_0/{\cal Z}$, 
$w_1^{\sigma}={\cal Z}_1^{\sigma}/{\cal Z}$, and ${\cal Z}={\cal Z}_0+
{\cal Z}_1^{\uparrow}+{\cal Z}_1^{\downarrow}$, with
\begin{eqnarray}
{\cal Z}_0&=&\prod_{n=-\infty}^{\infty}Z_n^{\uparrow}Z_n^{\downarrow},\cr
{\cal Z}_1^{\sigma}&=&\exp[-\beta(E_f+U-\mu-\frac{1}{2}g\mu_B\sigma h)]
\prod_{n=-\infty}^{\infty}(Z_n^{\uparrow}-U)(Z_n^{\downarrow}-U),
\label{eq: wdefs}
\end{eqnarray}
where $Z_n^{\sigma}:=[G_{\sigma}^0(i\omega_n)]^{-1}$.

The algorithm for finding the self-consistent solution to these equations
is the same as was used in the Hubbard model\cite{jarrell}: (i) begin
with the self energy set equal to zero; determine the local Green's
function from Eq.~(\ref{eq: gloc}); (iii) determine the site-excluded
Green's function from Eq.~(\ref{eq: g0def}) and the f-electron occupancies
from Eq.~(\ref{eq: wdefs}); (iv) determine the new local Green's function from
Eq.~(\ref{eq: gloc2}) and the new self energy from Eq.~(\ref{eq: self}).
This new self energy is plugged back into step (ii) and the process is iterated
until it converges (which typically takes less than 100 iterations for
convergence to one part in $10^6$).

In addition to finding the Green's functions in a magnetic field, we can also
determine the susceptibilities to both charge-density-wave order and 
spin-density-wave order in zero magnetic field.  The derivation of these
quantities is similar to that of Brandt and Mielsch\cite{brandt.1989},
and we simply include details not presented before and summarize the 
generalizations needed when the electrons have spin.

We begin by adding an ordering field $\sum_j h_j^{\sigma}d_{j\sigma}^{\dagger}
d_{j\sigma}$ to the Hamiltonian and examining the linear response in the
limit as $h_j^{\sigma}\rightarrow 0$.  The real-space conduction-electron
correlation functions can be expressed in terms of the Green's functions
via
\begin{equation}
\langle(n_{i\sigma}^d-\langle n_{i\sigma}^d\rangle )(n_{j\sigma^{\prime}}^d-
\langle n_{j\sigma^{\prime}}^d\rangle )\rangle =-T^2\sum_n
\frac{dG_{n\sigma}^{ii}}{dh_j^{\sigma^{\prime}}}.
\label{eq: correlation}
\end{equation}
Following the standard techniques,\cite{brandt.1989} one can express the
correlation functions in terms of charge-density-wave (CDW) ($\tilde\chi$)
and spin-density-wave (SDW) ($\chi$) susceptibilities in momentum space
\begin{eqnarray}
\tilde\chi_n^{dd}(q)&=&\chi_n^{dd0}(q)-T\sum_m\chi_n^{dd0}(q)
\tilde\Gamma_{nm}^{dd}\tilde\chi_m^{dd}(q),\cr
\chi_n^{dd}(q)&=&\chi_n^{dd0}(q)-T\sum_m\chi_n^{dd0}(q)
\Gamma_{nm}^{dd}\chi_m^{dd}(q),
\label{eq: dyson}
\end{eqnarray}
where the susceptibility is found by summing over Matsubara frequencies
[$\chi^{dd}(q)=T\sum_n\chi_n^{dd}(q)$], $\chi_n^{dd0}(q)$ is the
bare particle-hole susceptibility,
\begin{eqnarray}
\chi_n^{dd0}(q)&=&-T\sum_{k} G_{n}(k+q)G_{n}(k),\cr
&=&-\int_{-\infty}^{\infty}
dy \frac{\rho(y)}{i\omega_n+\mu-\Sigma_n-y}\int_{-\infty}^{\infty}dz\frac
{\rho(z)}{i\omega_n+\mu-\Sigma_n-X(q)y-z\sqrt{1-X^2(q)}},
\label{eq: bare_susc_def}
\end{eqnarray}
where all of the wavevector dependence  is included in the term
$X(q)=\sum_{j=1}^d (\cos q_j)/d$
(the Green's function and self energy in zero magnetic field have no spin
dependence), and $\tilde\Gamma_{nm}^{dd}$ ($\Gamma_{nm}^{dd}$) are
the irreducible CDW (SDW) vertex functions,
\begin{equation}
\tilde\Gamma_{nm}^{dd}=:\frac{1}{T}\left [ \frac{d\Sigma_{n\uparrow}}
{dG_{m\uparrow}}+\frac{d\Sigma_{n\uparrow}}{dG_{m\downarrow}}\right ] ,\ \ 
\Gamma_{nm}^{dd}=:\frac{1}{T}\left [ \frac{d\Sigma_{n\uparrow}}
{dG_{m\uparrow}}-\frac{d\Sigma_{n\uparrow}}{dG_{m\downarrow}}\right ] .
\label{eq: vertex_def}
\end{equation}
Now the self energy can be expressed as an explicit function of the
Green's function (of the same Matsubara frequency) and the f-electron filling
[by substituting Eq.~(\ref{eq: g0def}) into Eq.~(\ref{eq: gloc2}) and solving 
the resulting quadratic equation for $\Sigma_n$]
\begin{equation}
\Sigma_{n\sigma}=-\frac{1}{2G_{n\sigma}}+\frac{U}{2}\pm \frac{1}{2G_{n\sigma}}
\sqrt{1-2(1-2w_1)UG_{n\sigma}+U^2G_{n\sigma}^2},
\label{eq: sigma_sqrt}
\end{equation}
with $w_1=w_1^{\uparrow}+w_1^{\downarrow}$ the total f-electron concentration.
The difficulty in calculating the derivatives (to find the irreducible vertices)
arises from the fact that the f-electron filling $w_1$ is an explicit
function of all of the Green's functions.

The CDW vertex becomes
\begin{equation}
\tilde\Gamma_{nm}^{dd}=\frac{1}{T}\left ( \frac{\partial\Sigma_{n\uparrow}}
{\partial G_{n\uparrow}}\right )_{w_1}\delta_{mn}
+\frac{1}{T}\left ( \frac{\partial\Sigma_{n\uparrow}}{\partial w_1}
\right )_{G_{n\uparrow}} \left [ \frac{\partial w_1}{\partial G_{m\uparrow}}
+ \frac{\partial w_1}{\partial G_{m\downarrow}}\right ].
\label{eq: cdw_vert}
\end{equation}
In the zero-magnetic-field case, the two derivatives in the square brackets
are equal to each other, which simplifies the analysis.
Substituting into the Dyson equation Eq.~(\ref{eq: dyson}) then yields
\begin{equation}
\tilde\chi_n^{dd}(q)=\chi_n^{dd0}(q)\frac{1-\left (
\frac{\partial\Sigma_{n\uparrow}}{\partial w_1} \right )_{G_{n\uparrow}}
\gamma (q)}{1+\chi_n^{dd0}(q)\left ( \frac{\partial\Sigma_{n\uparrow}}
{\partial G_{n\uparrow}}\right )_{w_1}},
\label{eq: chi_middle}
\end{equation}
with the function $\gamma(q)$ defined by
\begin{equation}
\gamma(q)=:\sum_n \tilde\chi_n^{dd}(q)\left [ \frac{\partial w_1}
{\partial G_{m\uparrow}} + \frac{\partial w_1}{\partial G_{m\downarrow}}
\right ].
\label{eq: gamma_def}
\end{equation}
Multiplying Eq.~(\ref{eq: chi_middle}) by $\partial w_1/\partial G_{n\uparrow}
+\partial w_1/\partial G_{n\downarrow}$ and summing over $n$ yields an equation
for $\gamma(q)$.  Brandt and Mielsch show how to massage that formula into
a form where the individual derivatives can be explicitly calculated.  The
algebra is somewhat long, and entirely contained in their work.  The final 
result for $\gamma(q)$ is
\begin{equation}
\gamma(q)=\frac{\sum_n\frac{\left ( \frac{\partial w_1}{\partial Z_{n\uparrow}}
+\frac{\partial w_1}{\partial Z_{n\downarrow}}\right )\left [ 1-G_n^2
\left (\frac{\partial \Sigma_n}{\partial G_n}\right )_{w_1}\right ]}
{1+G_n\eta_n(q)-G_n^2
\left (\frac{\partial \Sigma_n}{\partial G_n}\right )_{w_1}} }
{1-\sum_n\frac{G_n\eta_n(q)\left ( \frac{\partial \Sigma_n}{\partial w_1}
\right )_{G_n}\left ( \frac{\partial w_1}{\partial Z_{n\uparrow}}
+\frac{\partial w_1}{\partial Z_{n\downarrow}}\right )}
{1+G_n\eta_n(q)-G_n^2
\left (\frac{\partial \Sigma_n}{\partial G_n}\right )_{w_1}} },
\label{eq: gamma_final}
\end{equation}
with $\eta(q)$ defined by 
\begin{equation}
\eta_n(q)=:G_n\left [ -\frac{1}{G_n^2}-\frac{1}{\chi_n^{dd0}(q)}\right ].
\label{eta_def}
\end{equation}
Finally, the CDW susceptibility is computed from
\begin{equation}
\tilde\chi^{dd}(q)=-T\sum_n\frac{[1-\gamma(q)\left ( \frac{\partial \Sigma_n}
{\partial w_1} \right )_{G_n}]G_n^2}
{1+G_n\eta_n(q)-G_n^2
\left (\frac{\partial \Sigma_n}{\partial G_n}\right )_{w_1}} .
\label{eq: chicdw_final}
\end{equation}
This susceptibility diverges when $\gamma(q)$ diverges, which occurs when
the denominator of Eq.~(\ref{eq: gamma_final}) vanishes.  This yields the same
result as Brandt and Mielsch for the spinless case, except for an additional
factor of 2 multiplying the sums in the numerator and the
denominator arising from the two derivatives of $w_1$ (which are both equal).  

Each of the derivatives appearing in Eqs.~(\ref{eq: gamma_final}) and 
(\ref{eq: chicdw_final}) can be directly calculated.  A straightforward 
differentiation and simplification [employing the quadratic equation 
whose solution gave Eq.~(\ref{eq: sigma_sqrt})] yields
\begin{equation}
\left ( \frac{\partial w_1}{\partial Z_{n\uparrow}}
+\frac{\partial w_1}{\partial Z_{n\downarrow}}\right )=
\frac{2w_1(1-w_1)UG_n^2}{(1+G_n\Sigma_n)(1+G_n[\Sigma_n-U])},
\label{eq: deriv_1}
\end{equation}
\begin{equation}
1-G_n^2 \left (\frac{\partial \Sigma_n}{\partial G_n}\right )_{w_1}=
\frac{(1+G_n\Sigma_n)(1+G_n[\Sigma_n-U])}{1+G_n(2\Sigma_n-U)},
\label{eq: deriv_2}
\end{equation}
and
\begin{equation}
G_n^2\left ( \frac{\partial \Sigma_n}{\partial w_1} \right )_{G_n}=
\frac{UG_n^2}{1+G_n(2\Sigma_n-U)}.
\label{eq: deriv_3}
\end{equation}
For the cases examined here, we never found any divergences of the CDW
susceptibility.

We must follow a similar procedure to find the SDW susceptibility, but
the algebra simplifies tremendously in zero external field, because the
irreducible SDW vertex becomes
\begin{equation}
\Gamma_{nm}^{dd}=\frac{1}{T}\left ( \frac{\partial\Sigma_{n\uparrow}}
{\partial G_{n\uparrow}}\right )_{w_1}\delta_{mn},
\label{eq: sdw_vert}
\end{equation}
since the derivatives with respect to $w_1$ are equal and hence cancel.
The SDW susceptibility then assumes the simple form
\begin{equation}
\chi^{dd}(q)=-T\sum_n\frac{G_n^2}{1+G_n\eta_n(q)-G_n^2
\left (\frac{\partial \Sigma_n}{\partial G_n}\right )_{w_1}} ,
\label{eq: chi_sdw_final}
\end{equation}
which is never expected to diverge at any finite temperature.

Brandt and Mielsch also show how to calculate the mixed susceptibilities
that correlate the d-electron charge (or spin) with the f-electron charge
(or spin).  The idea is that derivatives of $w_{1i}^{\sigma}$ with respect to
the ordering field $h_j^{\sigma^{\prime}}$ produce the mixed susceptibilities,
so a direct calculation of the $dd$-susceptibility involves total derivatives of
the self energy with respect to the Green's functions, which can be expressed
in terms of the irreducible vertex (as done above) to find the susceptibility,
or can be expressed as partial derivatives with respect to the Green's function
and the $f$-electron filling (since the self energy is a function of $G_n$
and $w_1$). Hence one can solve directly for the mixed susceptibilities
(the details appear in Brandt and Mielsch's work).  The final result for the
mixed CDW susceptibility is
\begin{equation}
\tilde\chi^{df}(q)=\frac{\left [ 1+G_n\eta_n(q)-G_n^2
\left ( \frac{\partial \Sigma_n}{\partial G_n}\right )_{w_1} \right ] 
\tilde\chi_n^{dd}
(q)+G_n^2}{G_n^2\left (\frac{\partial \Sigma_n}{\partial w_1}\right )_{G_n}}
=\gamma(q),
\label{eq: df_cdw}
\end{equation}
in which the Matsubara-frequency dependence actually drops out of the 
expression [after using Eq.~(\ref{eq: chi_middle})], and it is equal to 
$\gamma(q)$.  Similarly, we find
the mixed SDW susceptibility vanishes $\chi^{df}(q)=0$.

Now, we could have also calculated the mixed susceptibility by adding an
ordering field for the $f$-electrons, and differentiating the $d$-electron
Green's function with respect to the $f$-electron ordering field.  The results
of each of these calculations must be the same, since the $df$-susceptibility
is equal to the $fd$-susceptibility.  However, if we write out the expression
for the $fd$-susceptibility explicitly in terms of the derivatives with 
respect to the $f$-electron-ordering field, we find
that it includes a dependence on the $ff$-susceptibility.  Hence the 
$ff$-susceptibility can be determined from the $df$-susceptibility found above. 
In the CDW channel, one finds that the $ff$-susceptibility satisfies
\begin{equation}
\tilde\chi^{ff}(q)=\frac{1}{2T}\frac{\tilde\chi^{df}(q)}
{\sum_n\frac{G_n^2\left [ \left (\frac{\partial\Sigma_n}{\partial w_{1\uparrow}}
\right )_{G_n}+ \left (\frac{\partial\Sigma_n}{\partial w_{1\downarrow}}
\right )_{G_n}\right ]}
{1+G_n\eta_n(q)-G_n^2\left (\frac{\partial \Sigma_n}{\partial G_n}\right )_{w_1}
}}.
\label{eq: ff_cdw}
\end{equation}
This result can be further simplified following the same steps as above,
to rewrite the derivatives in terms of explicitly calculable quantities.
After a significant amount of algebra, similar to what was done for
the $dd$-susceptibility in the CDW channel, we find
\begin{equation}
\tilde\chi^{ff}(q)=\frac{\frac{1}{T}w_1(1-w_1)}
{1-\sum_n\frac{G_n\eta_n(q)\left (\frac{\partial \Sigma_n}{\partial w_1}
\right )_{G_n}\left (\frac{\partial w_1}{Z_{n\uparrow}}+\frac{\partial w_1}
{Z_{n\downarrow}}\right )}
{1+G_n\eta_n(q)-G_n^2\left (\frac{\partial \Sigma_n}{G_n}\right )_{w_1}}},
\label{eq: ff_cdw_final}
\end{equation}
which has the same denominator as the $dd$- and $df$-susceptibilities.
The CDW transition temperature, is thereby determined by the condition that
this denominator vanishes.

A similar analysis for the SDW channel does not yield any useful
information because $\chi^{df}(q)=0$.  Instead, we can directly compute
the uniform ($q=0$) $ff$-susceptibility for SDW order by differentiating
the fillings $w_1^{\sigma}$ with respect to the $f$-levels $E_f^{\sigma}$.
The calculation is simplified by the fact that the factors $Z_n^{\sigma}$
depend only on the total $f$-electron filling $w_1$ which does not change
in a magnetic field (only the relative fillings $w_1^{\uparrow}-
w_1^{\downarrow}$ change), and the final result is
\begin{equation}
\chi^{ff}(0)=\frac{w_1}{2T},
\label{eq: ff_sdw_final}
\end{equation}
which is in the Curie-Weiss form.

\section{Results}

We present the magnetic response of our numerical solutions for a variety of
different cases in Figure 1. Each figure corresponds to a different value of 
total electron concentration ($n=n_d+n_f$), 
with ten values of $E_f/t^{*}$ ranging from 0 to $4.5$ in steps 
of $0.5$. Fig.~1(a) is a typical high-density result. There are 2.5 electrons 
per impurity site, implying $n_f\neq 0$ at all temperatures, leading to a
Curie contribution to the susceptibility at small $T$. In this regime, the
results are rather insensitive to $U$ and $E_f$ 
($U=10t^{*}$ here). Fig.~1(b) is the $%
n=2.1$ case. We chose $U=t^{*}$ here, and the results have a stronger dependence
on $E_f$, showing a downturn at moderate $T$, before the Curie-law
divergence sets in for $T\rightarrow 0$. Neither of these cases have a 
charge-transfer transition, rather they always display a Curie-like
divergence as $T\rightarrow 0$.  As $n$ is decreased to 2 and
beyond, it is no longer necessary for there to be any $f$ electrons remaining at
$T=0$, and $\chi ^{ff}$ can vanish in that limit. In Fig.~1(c) we plot $\chi
^{ff}$ for $n=2$ and $U=2t^{*}$, showing an asymmetric peak that increases
in size and sharpens as the turnover temperature
$T_V$ decreases, and which still has a weaker Curie
upturn at the lowest temperatures, because $n_f\ne 0$ as $T\rightarrow 0$.
Here, the results depend on $E_f$ and $U$, with the peak
lowering in magnitude and broadening as $U$ increases, and the
low-temperature upturn becoming more prominent. The case $n=1.9$ is shown
in Fig.~1(d) for $U=10t^{*}$. Once again we have a dependence on
$E_f$ and $U$, but the low-temperature upturn has disappeared. In some
cases $\chi ^{ff}$ has a double-hump structure. Finally the low-density
regime is plotted in Fig.~1(e) ($n=0.5$, $U=10t^{*}$). Here, as in the
high-density limit, the results are insensitive to $U$.  The reason why
the results do not depend sensitively on $U$ for both the high and low-density
limits is that when $n_f$ is small, then the Falicov-Kimball model looks
like a noninteracting free band, because there are no localized electrons
for the conduction electrons to scatter off of.  Similarly, in the high
density regime, as $n_f$ approaches 1, the model also becomes noninteracting,
since there is an $f$-electron at every site, and the conduction band is simply
shifted in energy by $U$, which does not affect any of the qualitative
physics.

The model also exhibits a metamagnetic transition, for large enough magnetic
fields. This is shown in Figure 2 for one case: $n=2$, $E_f=t^{*}$, $%
U=2t^{*}$, $g=4.5$, and eight temperatures from $0.05t^{*}$ up to $6.4t^{*}$. 
The numerical analysis shows that at $T=0$, the Zeeman energy at the transition
is of the order of $k_BT_V$. The metamagnetic field,
$H_V$, decreases at higher temperatures but 
the transition becomes smoother. The metamagnetism disappears as $T$ is
increased beyond $T_V$, because the $f$-electrons are already occupied in the
zero-field limit.

This anomalous behavior can be understood in terms of simple thermodynamic 
considerations. For $E_f>\mu $ and $n<2$, the nonmagnetic empty
$f$-orbital is energetically more favorable than the magnetic state, 
so the ground state has no $f$-electrons. However, at high $T$, the large 
magnetic 
entropy of the $f$-electron spins favors the magnetic state. The susceptibility
of these $f$-electrons grows rapidly as the temperature is reduced until, close
to $T_V$, the entropy gain is insufficient to compensate the energy loss and
the $f$-electrons disappear to form a nonmagnetic state. Thus, $\chi ^{ff}$
drops rapidly, which is the origin of the sharp asymmetric peak in the 
$\chi (T)$ data. 

\section{Application to real materials}

The insulating state of NiI$_2$ is best described by a hole picture.
The localized holes lie in the Ni $d$-bands, and there is one hole per site.
The Iodine $p$-band is full, so there are no conduction holes.  Hence the
total hole concentration is 1 and in the regime where the pressure is tuned
to lie just above the metallization pressure at $T=0$,
we expect the susceptibility to look
qualitatively like the low-density regime of Fig.~1(e) (this is because the
only difference between the hole picture and the electron picture is that
the $f$-level energy changes sign).  So, we expect an interesting anomalous
magnetic response of NiI$_2$ when the pressure is sufficiently high that
the system is just on the metallic side of the metal-insulator transition 
point.  In this case, there are few localized holes at low temperature, but they
will be thermally excited in large 
numbers as the temperature is raised.  The susceptibility should be
quite small at low temperature, and then rise dramatically as the 
metal-insulator transition point is passed (before turning over again
at higher temperature). Also, we predict that NiI$_2$ should exhibit
metamagnetism.  Since new techniques allow
one to inductively
measure the magnetic response of samples in a diamond-anvil cell,
\cite{hemley} without the need of attaching any wires within the cell,
it is possible that one could use this anomalous magnetic response to
accurately measure the phase diagram of NiI$_2$.  To our knowledge no 
measurements of the magnetic response of NiI$_2$ under pressure have
yet been carried out.  Only the N\'eel temperature in the insulating
phase was measured with the M\"ossbauer effect.\cite{nii2}

For the  Yb-based VF compounds,
the mapping of this system to an effective 
Falicov-Kimball model is more complicated than in 
NiI$_2$, and more controversial, so we begin
by motivating how such a mapping can be made, and provide experimental evidence
that supports this novel point of view.
We notice first that there is no simple relation between the change in the
lattice parameter (that is, the change in Yb-In hybridization)
and the onset of the anomalies. In Yb$_{1-x}$Y$_x$InCu$_4$ 
an increase of $a(x)$ is accompanied by a decrease of $T_V$ and $H_V$. 
In YbIn$_{1-x}$Ag$_x$Cu$_4$, $a(x)$ does not change when In ions are replaced 
by smaller Ag ions (for $x\leq 0.4$) while $T_V$ and $H_V$ increase and the 
anomalies become less sharp\cite{yoshimura.90,sarrao.96}. 
In stoichiometric YbInCu$_4$, $T_V$ and $H_V$ are 
enhanced by thermal treatment, which introduces disorder between Yb
and In sites but does not change $a$ (or the Yb-In hybridization). We notice 
also that systems which
are similar for $T\gg T_V$ ($H\gg H_V$) and $T\ll T_V$ ($H\ll H_V$)
can show substantial variation in the onset and the shape of the anomalies
at $T_V$ and $H_V$. 
%
%
%
To account for such a behavior we assume that the effective 
f-state is in the proximity of the chemical potential, $\mu $, 
so that correctly placed Yb ions are in the VF state. Hence, 
the important parameter is the renormalized position of the f-level, 
$E_f-\mu $, which provides the characteristic energy scale (typically, 
about 100 K). 

The assumption that the majority of Yb ions are in the VF state for 
$T\gg T_V$ is supported by the unusual deviation from Vegard's law
found in YbIn$_{1-x}$Ag$_x$Cu$_4$ for $x\leq 0.4$ at room
temperature.\cite{sarrao.96}
Here, the replacement of In ions by smaller Ag ions 
does not modify  $a(x)$, since in a VF compound
the Yb ions reduce their average valence and preserve
the volume of the unit cell. The assumption that $E_f-\mu $ increases only 
slightly with $x$, explains the weak concentration dependence of the
high-temperature susceptibility and the high-field saturation magnetization%
\cite{yoshimura.90,sarrao.96}. 
For $x\leq 0.4$, $E_f$ is pushed off $\mu $, which results
in a HF compound with a stable f-shell.\cite{yoshimura.90,sarrao.96}

To explain the anomalies at $T_V$, we assume that the magnetic response of these
Yb compounds has its origin in two physically distinct components. 
The first one, $\chi _{VF}$, is due to the hybridized (correctly placed)
Yb ions, and is large (on an absolute scale), isotropic, and sample independent.
The second component of the susceptibility, $\chi ^{ff}$, arises from the 
unhybridized Yb ions which switch at $T_V$ (or $H_V$) from the magnetic 
$3+$ to the nonmagnetic $2+$ configuration. 
Hence, $\chi ^{ff}$ vanishes below $T_V$ and is smaller than $%
\chi _{VF}$ at high temperatures, but dominates $\chi (T)$ near $T_V$; it is
strongly sample dependent. At low temperatures and high fields these
unhybridized states lead to a metamagnetic transition with a small energy
difference between the low-field and the high-field states.

The lattice of YbInCu$_4$ unit cells separates into two distinct sublattices:
the sublattice of unhybridized Yb ions (localized f-electrons) and the
sublattice of hybridized Yb ions.  Conduction electrons from the In ions
move (via nearest-neighbor hopping between unit cells of the full lattice)
between any two unit cells of the unhybridized-Yb-ion sublattice.  Hence,
if we focus only on the sublattice degrees of freedom, we can model the d-band
by a {\it single effective band} that couples any two sites of the sublattice
with a {\it random} hopping matrix element (that depends on the growth and
thermal treatment of the sample). These $f$ and $d$ states
have a common chemical potential ($\mu $) and interact by a Coulomb
repulsion ($U$) when they occupy the same unit
cell. In addition, both the $d$ and $f$ particles carry a spin
label $\sigma $ and the $d$-level can accommodate $2$ electrons (or holes)
of the opposite spin. The occupancy of the $f$-level is restricted to $%
n_f\leq 1$ because of the large Coulomb repulsion ($U_{ff}\sim \infty $) of
the $f$-particles of opposite spins. To discuss the Yb-based VF compounds we
use the hole-picture, in which $E_f<\mu $ and the total number of holes at
the ill-placed sites is restricted to $n_h=n_d^h+n_f^h\leq 3$. (In the
electron-picture, one would have $E_f>\mu $ and would restrict the total
number of electrons to $n_e=n_d^e+n_f^e\leq 3$.) The magnetic field $h$
couples to the $f$ and $d$ states but with different $g$-factors ($g=4.5$
for the $f$-holes). This picture is thus described by the spin one-half
Falicov-Kimball model.
In the limit $N\rightarrow \infty $ \cite{kotliar}, and $N/N_l\ll 1$, our
choice for $t_{ij}$ maps this problem onto the infinite-dimensional (local)
one, which allows the magnetic susceptibility to be evaluated exactly using
the methods described above. 

The hole filling for these Yb-based VF compounds lies near $n\approx 2$.
The exact solution reproduces well the overall behavior of the experimental
data. Figs.(1c)-(1d) capture most of the features 
shown\cite{felner.86,yoshimura.88,kindler.94,sarrao.96} by $\chi(T)$, such as 
an asymmetric
peak at $T_V$ that increases in magnitude and sharpens as $T_V\rightarrow 0$%
, while Fig.(2) explains the magnetization\cite{yoshimura.88,katori.94} $M(T)$,
with a metamagnetic transition that smoothes out and then disappears as $T$
is increased. Various samples will have different numbers of Yb impurities
and will require different values of $t^*$, $E_f$, and $n^h$. Our analysis
shows that the lower the transition temperature, the more pronounced and
steeper the anomaly. For large fields, unhybridized Yb ions switch to a
magnetic configuration at much lower temperatures, than in the absence of
the field.

We emphasize that in the model proposed here, most of the Yb ions are in a 
hybridized (VF) state at all temperatures and that the anomalies around 
$T_V$ are {\it due to an entropy driven transition of a small number of
unhybridized Yb ions}, so that the bulk of the lattice is unchanged at $T_V$. 
On the other hand, in the commonly used 
hybridization model it is assumed that all the Yb ions switch at $T_V$ from 
a stable $3+$ configuration to a hybridized VF configuration, which makes 
the high-T and the low-T phases fundamentally different. 
Thus, the appropriate model for describing the Yb-based VF compounds 
can be determined by measuring the pressure dependence of 
$\Delta a/a$ across $T_V$. The Falicov-Kimball model predicts a weak pressure 
dependence because the ions involved in the transition are unhybridized, 
while the hybridization model predicts a strong pressure dependence 
because the hybridization rapidly increases with pressure.

\section{Conclusions}

In summary, we have exactly solved the spin-one-half Falicov-Kimball model
on an infinite-dimensional hypercubic lattice
in an external magnetic field.  We also examined the magnetic
response (as a function of temperature) in zero magnetic field.  The system
showed anomalous behavior due to the proximity of the $f$-electron states
(which have local moments) to the chemical potential, which allows their
occupancy to change dramatically as the temperature changes, because of
the entropy gain due to the magnetic moments.  Thus, the uniform magnetic
susceptibility shows an asymmetric peak at a characteristic temperature
$T_V$, which decays like a Curie-Weiss
law for high temperatures (because of the $f$-moments) and typically decays
exponentially rapidly for temperatures much lower than $T_V$ (because the
$f$-electrons are thermally excited across a gap).  The magnetization
(in an external magnetic field) shows metamagnetic behavior because of the
rapid switching of electrons from the $d$- to the $f$-states when $T<T_V$,
but shows a much smoother increase with magnetic field for temperatures
larger than $T_V$.

We applied this model to two candidate real materials.  The first is NiI$_2$,
which is known to undergo a charge-transfer metal-insulator transition as
a function of pressure, that is well described by the Falicov-Kimball model.
We propose that the metal-insulator phase diagram, and this anomalous
magnetic behavior can be easily measured using newly developed techniques
within a diamond-anvil cell.  The second are the
Yb-based VF compounds whose anomalous magnetic properties can be attributed 
to an entropy-driven transition of disordered Yb and explained by the 
Falicov-Kimball model with random hopping. The Falicov-Kimball-transition 
of unhybridized Yb ions 
and the VF behavior of hybridized Yb ions are both due to the proximity 
of the f-level to the chemical potential. The two inequivalent Yb sites appear 
in these compounds because of their characteristic crystal structure.
Thus, the large changes in magnetic, transport, and elastic properties 
at $T_V$ and $H_V$ are explained, and reconciled with small changes 
in the $f$-electron valence and $\Delta a$. This proposed 
alternative explanation for the magnetic response of the Yb-based VF
compounds is controversial, and differs from the more conventional 
description (which does not have any quantitative theoretical
description).  We propose high-pressure experiments to differentiate between
the two proposed theories.

\acknowledgments

We acknowledge useful conversations with I. Aviani, Z. Fisk, M. Jarrell, B.
L\"uthi, M. Miljak, and J. Sarrao. J.~K.~F. acknowledges the Donors of The
Petroleum Research Fund, administered by the American Chemical Society, for
partial support of the early stages of this
research (ACS-PRF No. 29623-GB6) and the Office of
Naval Research Young Investigator Program (N000149610828) for supporting the 
later stages of this work. The early stages of this project have also
been funded in part by the National Research Council under the Collaboration
in Basic Science and Engineering Program. The contents of this publication
do not necessarily reflect the views or policies of the NRC, nor does
mention of trade names, commercial products, or organizations imply
endorsement by the NRC.

$^{\dag}$ Permanent address: Institute of Physics, Zagreb, Croatia.

\begin{figure}[t]
\caption{Magnetic susceptibility of the $f$-electrons in the Falicov-Kimball
model. The different curves correspond to $E_f/t^*=0.0, 0.5,\ldots,4.5$ (in
general $E_f$ increases from top to bottom in these figures).
The different figures are: (a) $n=2.5$ and $U=10t^*$; (b) $n=2.1$ and 
$U=t^*$;
(c) $n=2.0$ and $U=2t^*$; (d) $n=1.9$ and $U=10t^*$; and (e) $n=0.5$ and 
$U=10t^*$.}
\end{figure}

\begin{figure}[t]
\caption{Total magnetization of the Falicov-Kimball model as a function
of magnetic field.  The different curves correspond to different temperatures:
$T/t^*=0.05,0.1,0.2,\ldots,3.2,6.4$ (the temperature increases from top to 
bottom
in the large-$h$ range).  The parameters are $n=2$, $E_f=t^*$, and $U=2t^*$.}
\end{figure}


\begin{figure}[tbp]
\epsfxsize=4.5in \epsffile{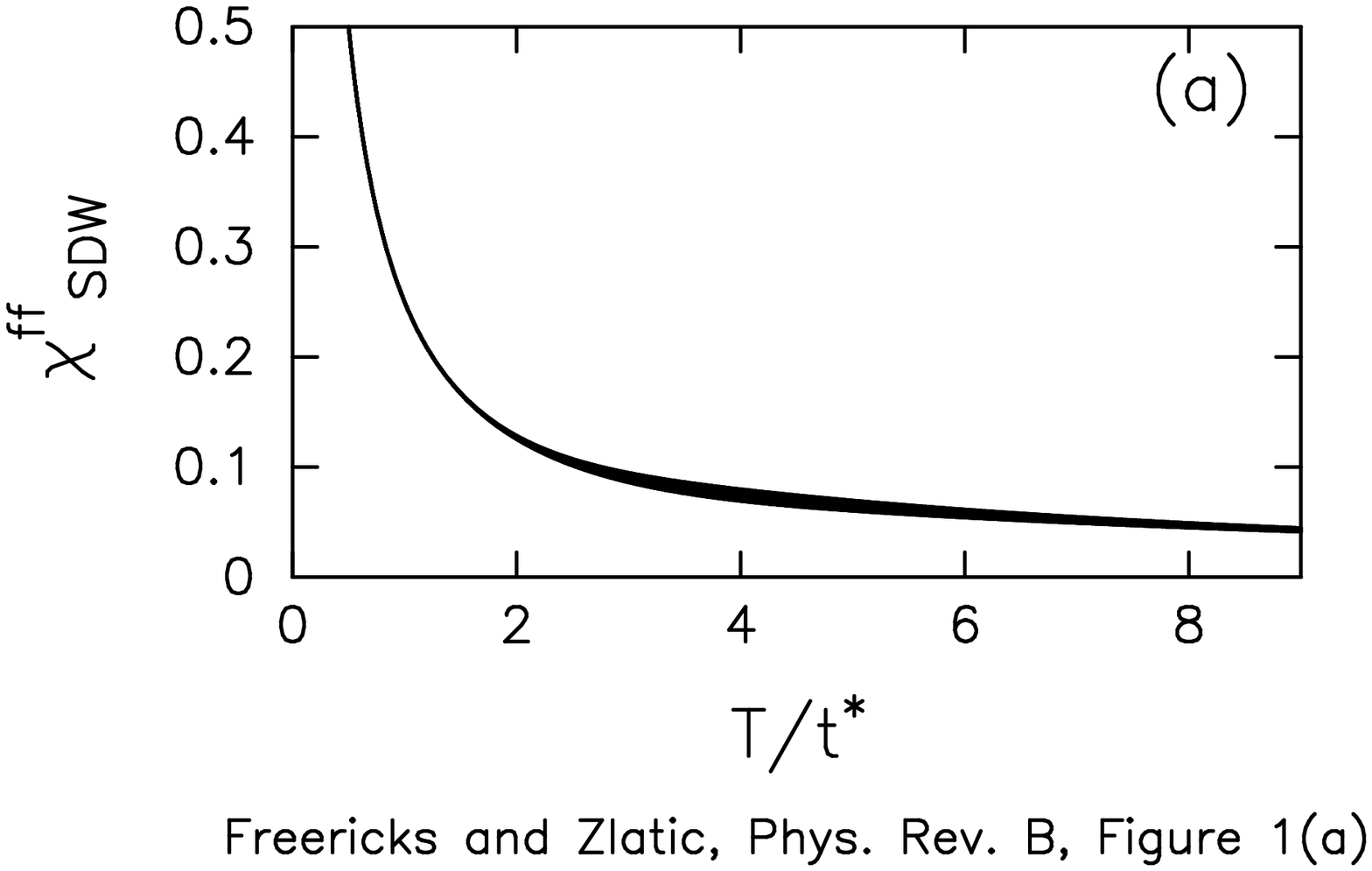}
\end{figure}

\begin{figure}[tbp]
\epsfxsize=4.5in \epsffile{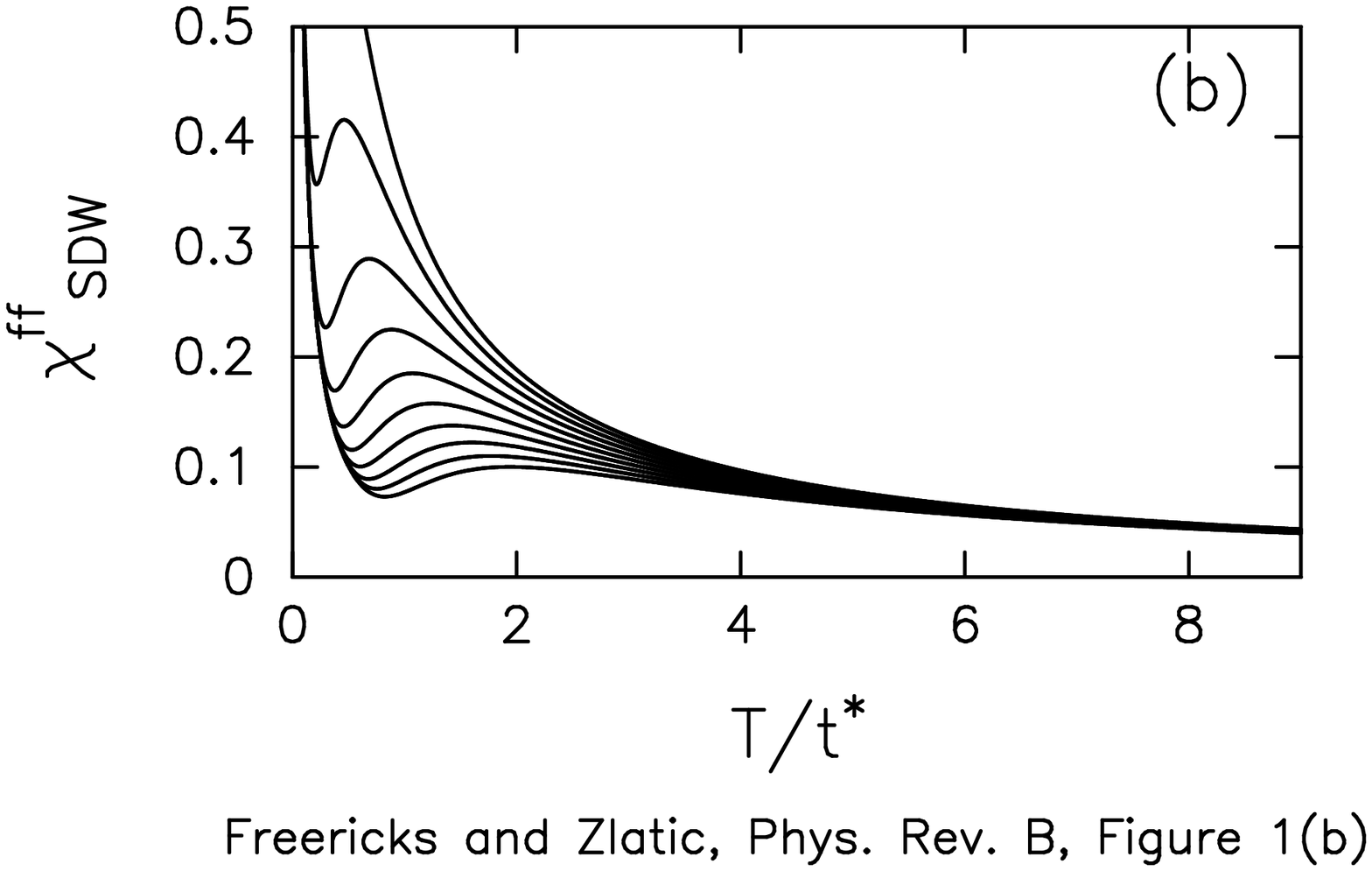}
\end{figure}

\begin{figure}[tbp]
\epsfxsize=4.5in \epsffile{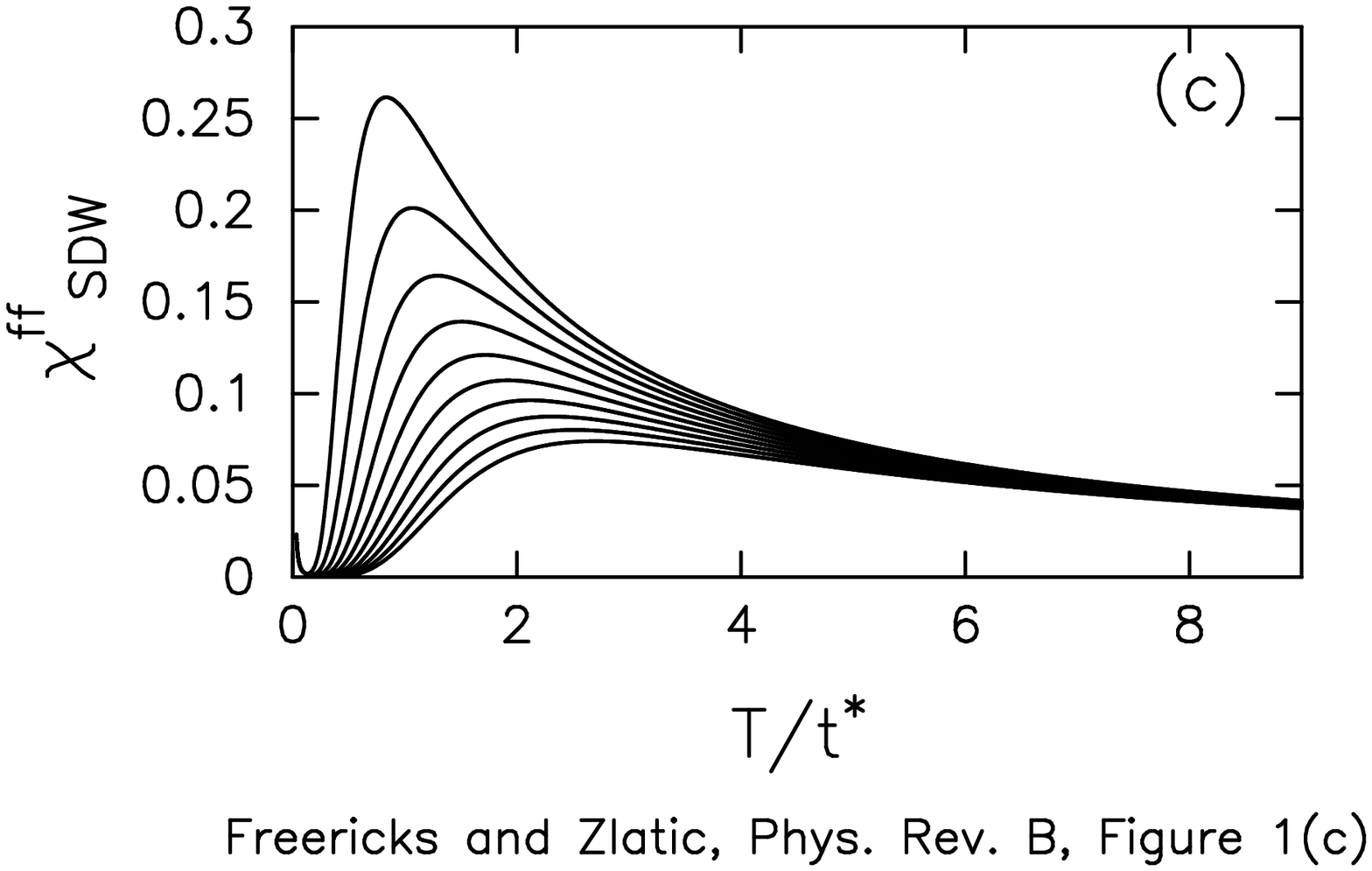}
\end{figure}

\begin{figure}[tbp]
\epsfxsize=4.5in \epsffile{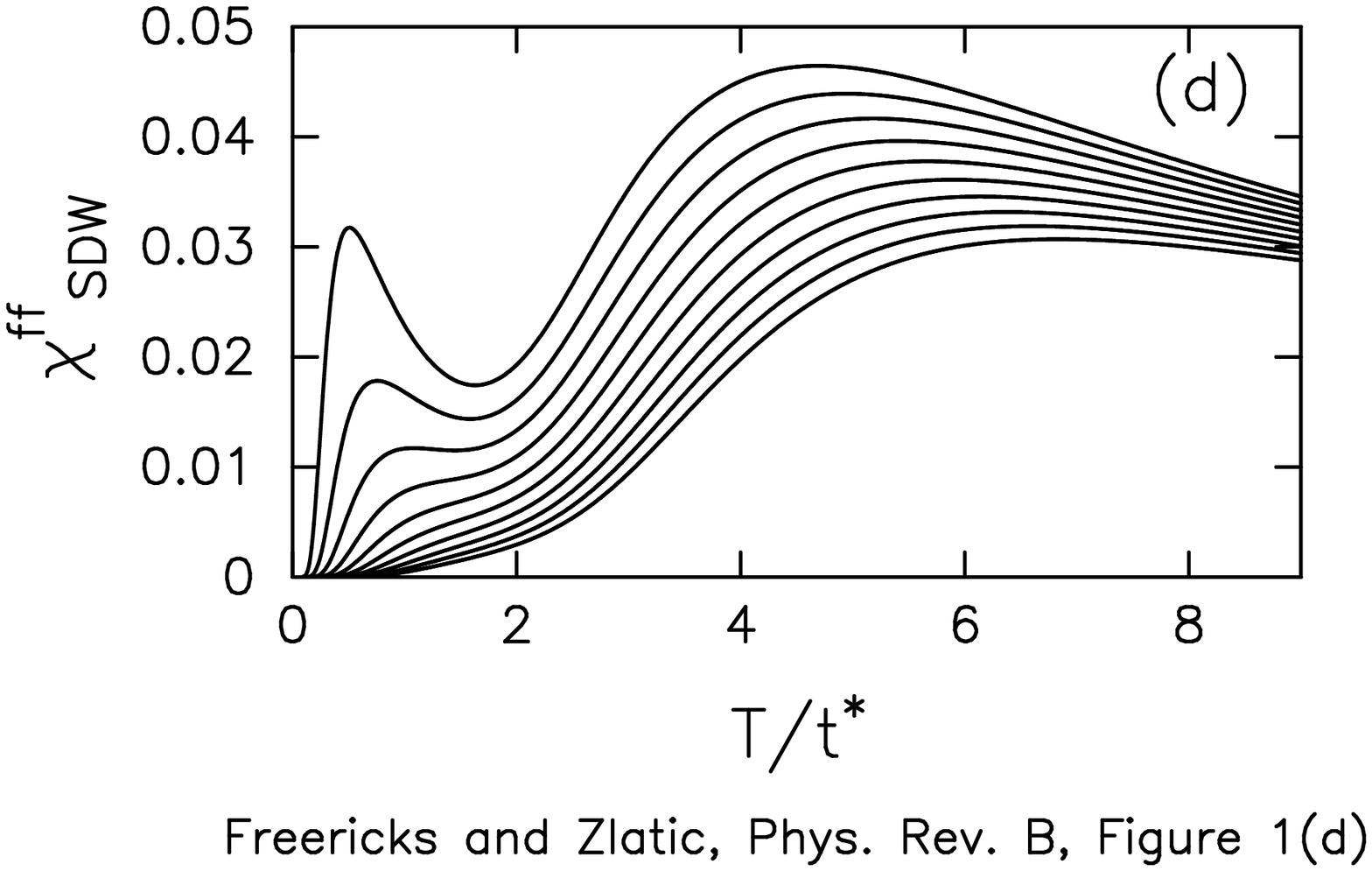}
\end{figure}

\begin{figure}[tbp]
\epsfxsize=4.5in \epsffile{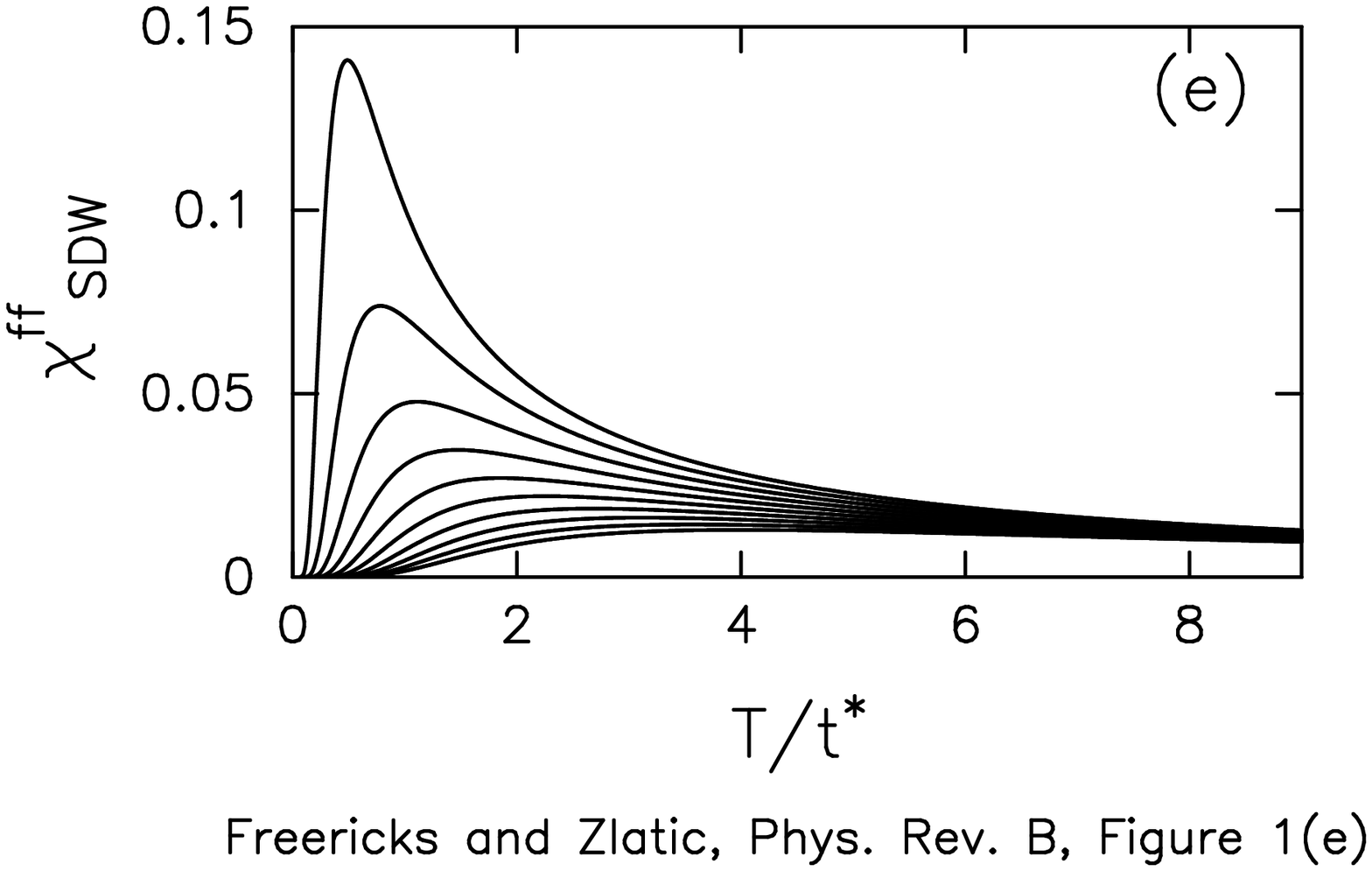}
\end{figure}

\begin{figure}[tbp]
\epsfxsize=4.5in \epsffile{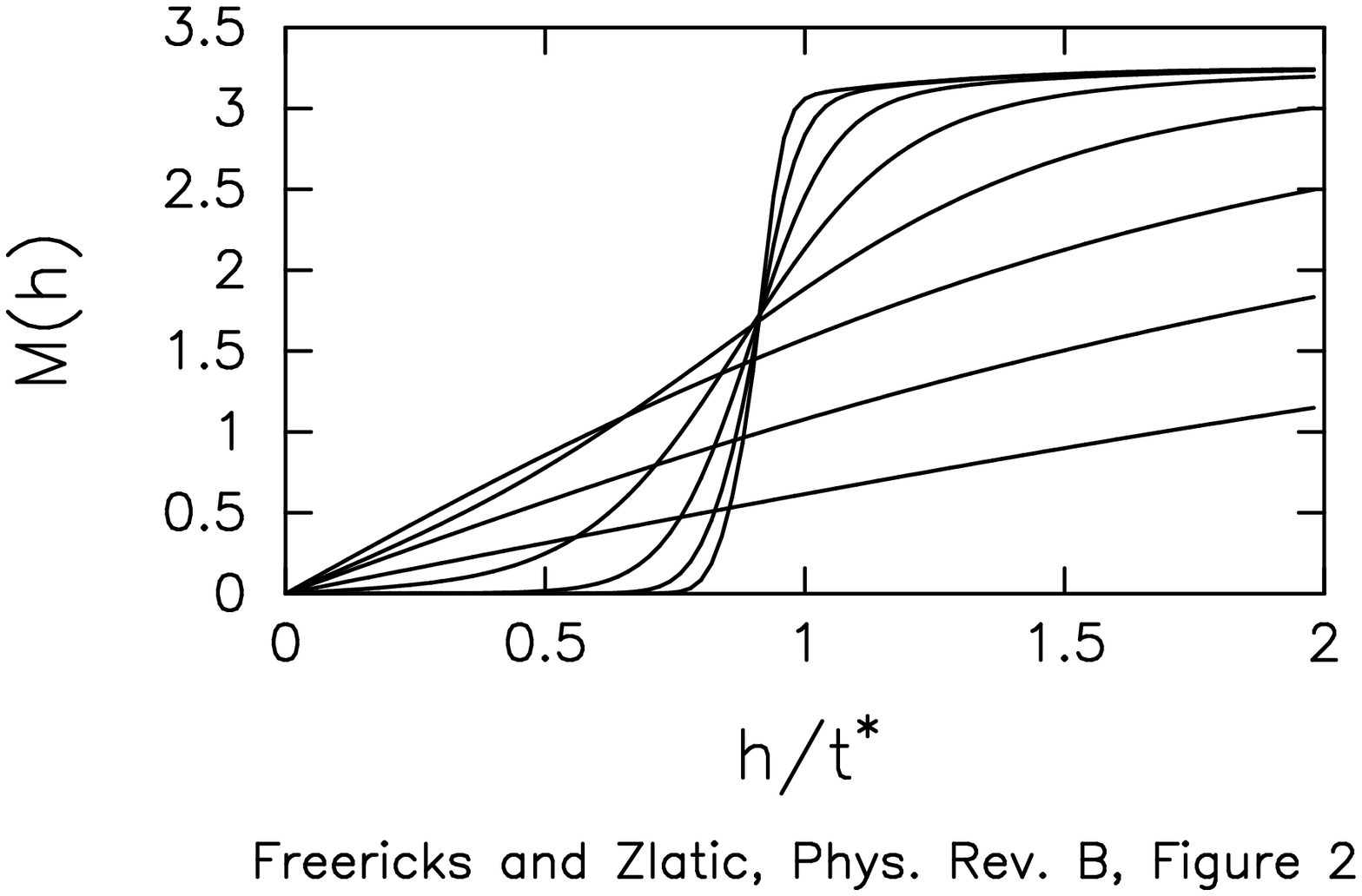}
\end{figure}


\begin{references}

\bibitem{falicov.69}  L. M. Falicov and J. C. Kimball, Phys. Rev. Lett. {\bf %
22}, 997 (1969); R. Ramirez, L. M. Falicov and J. C. Kimball, 
Phys. Rev. B {\bf 2}, 3383 (1970).

\bibitem{nii2}  M. P. Pasternak, R. D. Taylor, A. Chen, C. Meade, 
L. M. Falicov, A. Giesekus, R. Jeanloz, and P. Y. Yu, Phys. Rev. Lett. 
{\bf 65}, 790 (1990); J. K. Freericks and L. M. Falicov, Phys. Rev. 
{\bf 45}, 1896 (1992); A. L. Chen, P. Y. Yu, and R. D. Taylor, Phys. Rev. Lett. 
{\bf 71}, 4011 (1993).

\bibitem{felner.86}  I. Felner and I. Novik., Phys. Rev. B {\bf 33}, 617
(1986); I. Felner, et al., Phys. Rev. B {\bf 35}, 6956 (1987).

\bibitem{yoshimura.88}  K. Yoshimura, et al., Phys. Rev. Lett. {\bf 60}, 851
(1988); T. Shimizu et al., J. Phys. Soc. Japan {\bf 57}, 405 (1988).

\bibitem{kindler.94}  B. Kindler, et al., Phys. Rev. B {\bf 50}, 704 (1994).

\bibitem{kojima.90}  K. Kojima, et al., J. Phys. Soc. Japan, {\bf 59}, 792
(1990).

\bibitem{nakamura.90}  H. Nakamura, et al., J. Phys. Soc. Japan {\bf 59}, 28
(1990).

\bibitem{sarrao.96}  J.L. Sarrao et al., Physics B, {\bf 223\&224}, 366
(1996).

\bibitem{junhui.97}  H. Junhui, et al., J. Phys. Soc. Japan, {\bf 66}, 2481
(1997).

\bibitem{lawrence.96}  J. M. Lawrence et al., Phys. Rev. B {\bf 54}, 6011
(1996).


\bibitem{katori.94}  H.A. Katori, et al., Physica B, {\bf 201}, 159 (1994).

\bibitem{nonlinear} T. Portengen, T. Ostreich, and L. J. Sham, Phys. Rev. B
{\bf 54}, 17452 (1996); Phys. Rev. Lett. {\bf 76}, 3384 (1996).

\bibitem{farkasovsky} P. Farka\v{s}ovsk\'y, Phys. Rev. B {\bf 51},
1507 (1995); {\bf 52}, R5463 (1995); {\bf 54}, 11261 (1996); 
Z. Phys. B {\bf 104}, 147 (1997); {\bf 104} 553 (1997); M. Park and
J. Hong, unpublished.

\bibitem{metzner_vollhardt} W. Metzner and D. Vollhardt, Phys. Rev. Lett.
{\bf 62}, 324 (1989).

\bibitem{brandt.1989}  U. Brandt and C. Mielsch, Z. Phys. B {\bf 75}, 365
(1989); Z.Phys. B {\bf 79}, 295 (1990); U. Brandt, A Fledderjohann,
and G. H\"ulsenbeck, Z.
Phys. B {\bf 81}, 409 (1990); U. Brandt and A. Fledderjohann, Z. Phys. B
{\bf 87}, 111 (1992).

\bibitem{jarrell} M. Jarrell, Phys. Rev. Lett. {\bf 69}, 168 (1992).

\bibitem{hemley}  V. V. Struzhkin, Y. A. Timofeev, R. J. Hemley, and H.-K. Mao,
Phys. Rev. Lett. {\bf 79}, 4262 (1997).

\bibitem{yoshimura.90}  K. Yoshimura et al., J. Mag. Mag. Mat. {\bf 90\&91},
466 (1990).




\bibitem{pillmayr.92}  N. Pillmayr et al., J. Mag. Mag. Mat. {\bf 104\&107},
639 (1992).

\bibitem{dos}  We choose a Gaussian density of states rather than Wigner's
semicircular density of states in order to include the effects of Lifshitz's
tails. The theory is rather insensitive to the precise choice of $\rho
(\epsilon )$. For details on this choice, see M. Moshe, H. Neuberger, and B.
Shapiro, Phys. Rev. Lett. {\bf 73}, 1497 (1994); M. Kreynin and B. Shapiro,
Phys. Rev. Lett. {\bf 74}, 4122 (1995).

\bibitem{kotliar}  M. J. Rozenberg et al., Phys. Rev. B{\bf 49}, 10181
(1994).
\end{references}
\end{document}